\documentclass[aps,twocolumn,prl,showpacs]{revtex4}
\usepackage{graphicx}
\usepackage{amssymb}
\usepackage{amsmath}
\usepackage{color}
\begin{document}

\title{General conditions for a quantum adiabatic evolution}

\author{Daniel Comparat}
\email{Daniel.Comparat@lac.u-psud.fr}
\affiliation{Laboratoire Aim\'e Cotton\footnote{Laboratoire Aim\'{e} Cotton is
associated to Universit\'{e} Paris-Sud and belongs to F\'{e}d\'{e}ration de
Recherche Lumi\`{e}re Mati\`{e}re (LUMAT).}, Univ Paris-Sud 11, Campus d'Orsay B\^at. 505, 91405 Orsay, France}

\date{}
\begin{abstract}
The smallness of the variation rate of the hamiltonian matrix elements compared to the (square of the) energy spectrum gap   is usually believed to be the key parameter for a quantum adiabatic evolution. However it is only perturbatively  valid
for scaled timed hamiltonian and resonance processes as well as off resonance possible constructive St\"uckelberg interference effects violate this usual condition for general hamiltionian. More
general  adiabatic condition and exact bounds for adiabatic quantum evolution are derived and studied in the framework of a two-level system. The usual criterion is  restored for real two level hamiltonian with small number of monotonicity changes of the hamiltonian matrix elements and its derivative.

\end{abstract}


\maketitle

Adiabaticity is at the border between dynamics and statics. It has been introduced by Boltzmann in classical mechanics and by Born and Fock in 1928
in Quantum Mechanics \cite{Nakamura2002,Teufel2003}, extended  to the infinite dimensional setting by Kato (1950), studied as a
geometrical holonomy evolution by Berry (1984), finally extended
to degenerate cases (without gap condition) and to open quantum system more recently \cite{1998PhRvA..58.4300A,2005PhRvA..71a2331S}. 
The quantum adiabatic theorem 
is usually used to derive
approximate solutions of the Schr\"odinger equation and
is strongly related to the (semi-)classical limit $\hbar \rightarrow 0$ of quantum mechanics \cite{1984JPhA...17.1225B} and to the 	
	Minimal work principle \cite{2005PhRvE..71d6107A}  for the Hamiltonian $H(t)$. 
The principle is simple: if a quantum system is prepared in an eigenstate $| n(0) \rangle $ of a
``slowly'' varying Hamiltonian it remains (without taking into account of the phase evolution)  close to the instantaneous eigenstate $| n(t) \rangle $
of this Hamiltonian as time $t$ goes on. 
The applications
range from two-level systems (nuclear magnetic resonance, atomic laser transitions, Born-Oppenheimer molecular adiabatic coupling, collisional processes  ...) to quantum  algorithms \cite{2001Sci...292..472F}.

``Usual''
adiabatic conditions  are  (for all $t\in [0,T]$):
\begin{equation}
 \sum_{m \neq n} \frac{ 1 }{|\omega_{m n} (t)|}  \frac{|\dot H_{m  n} (t)| }{  |E_{m n}(t)| } = 
\sum_{m \neq n} \left| \frac{ \langle m (t) | \dot  n(t) \rangle }{\omega_{m n} (t)} \right | \ll 1 ,\label{adia_crit}
\end{equation}
 where  the dot designs the time derivative and $|m \rangle$ are the instantaneous eigenstates for the energy eigenvalue $E_m(t)$ with  $E_{m n}  = \hbar \omega_{m n} = E_m - E_n $ \footnote{We use the time derivative of $\langle m  | n \rangle $ and $\langle m | H | n \rangle $ leading,  for non degenerate case, to 
$
- \langle \dot m  | n \rangle = \langle m  | \dot n \rangle = -
 \frac{\langle m | \dot H | n \rangle }{E_m - E_n }. 
$
}.
Some confusion occurs recently \cite{2004PhRvL..93p0408M,2005PhRvL..95k0407T,2005PhRvA..71f3409C,2005quant.ph.10131D,2005PhLA..339..288T,2004quant.ph..5129P} because, this condition seems written for a general hamiltonian $H(t) $. However, it has been studied by many different techniques (see for instance \cite{2005math.ph..11067H,2006quant.ph..3175J}) but only for special types of hamiltonian such as time scaling one $H(t) = \hat H (t/T) $ \footnote{An important example is the 
interpolating hamiltonian
$H(t) =H_{\rm in} (1-t/T)  + H_{\rm fin} t/T $.
 $ H (t) = \hat H (s(t)) $ have also been considered  with
 a \emph{monotonic} function 
$s(t)\in [ 0,1 ]$ controlling locally the speed of the process.
When the timing $T$ is not an issue  $s = t/T$ is the simplest choice.}. 
 Furthermore, even for such a time scaled hamiltonian,  condition (\ref{adia_crit}) is not sufficient because it is only the leading order term
\cite{2006PhLA..353...11V,2006PhRvA..73d2104M}, in a time evolution $T$ perturbation point of view, and more accurate conditions are needed to prove adiabatic evolution \cite{2006quant.ph..3175J}. 
 
  The goal of this article is to 
 derive general quantum adiabatic conditions for general hamiltonian. 
We start our study on a two level system example in order to study some possible violation of the usual adiabatic conditions.
Afterwords, considering a more general  type of $N$ levels hamiltonians, we derive a  general criterion for adiabaticity.
 Finally, the study of the interference during multiple passages allows us to precise the  validity of  the usual adiabatic condition.
     
A quite general $2 \times 2$ hamiltonian matrix, written in the Pauli Matrix ($\vec \sigma$) basis, leads to the a spin $1/2$ form
$H =-\hbar \frac{\gamma}{2} \vec B.\vec \sigma$:
$$ H = -  \frac{\hbar \omega_0}{2}  \left( \begin{smallmatrix} \cos \theta  & \sin \theta e^{- i \varphi} \cr 
\sin \theta e^{i \varphi} & - \cos \theta
\end{smallmatrix} \right)  = -  \frac{\hbar }{2}\left( \begin{smallmatrix} \delta_0 + \omega_L & \Omega_0 e^{ i \int \omega_L } \cr 
\Omega_0 e^{- i \int \omega_L} & - \delta_0 - \omega_L
\end{smallmatrix} \right)
$$
where  $\omega_0 = \gamma B$ is the Larmor frequency,
$\vec B$ is a rotating  magnetic field with a polar angle $\theta$, an  azimuthal rotating angular frequency
$\dot \varphi=-\omega_L$.
Where the second form of the hamiltonian represents, in the rotating wave approximation (RWA), a two level system coupled to an external (laser with angular frequency $\omega_L$ for instance) field  which is frequency detuned by $\delta_0 = \omega_0 \cos \theta - \omega _L$ from the resonance and with a real Rabi frequency $\Omega_0 = \omega_0 \sin \theta$. For future developments we also
  define  $\Omega_L = \omega_L \sin \theta - i \dot \theta = |\Omega_L| e^{i \arg \Omega_L}  , \delta_L =  \omega_L \cos \theta - \omega_0 + \frac{d}{d t} \arg \Omega_L$ and
$\Omega_R= \sqrt{|\Omega_L|^2+\delta_L^2} $.
 The eigenvectors $e^{i \theta_\mp} |\mp \rangle$, corresponding to the eigenvalues $\mp \hbar \omega_0/2$, are given by the columns of $R_\theta= \left( \begin{smallmatrix}
e^{ -i \frac{ \varphi}{2}} \cos \frac{\theta}{2} e^{i \theta_-} & - e^{- i \frac{ \varphi}{2}} \sin \frac{\theta}{2} e^{i \theta_+}   \cr 
 e^{ i \frac{ \varphi}{2}} \sin \frac{\theta}{2} e^{i \theta_-}  & e^{ i \frac{ \varphi}{2}} \cos \frac{\theta}{2} e^{i \theta_+} 
\end{smallmatrix} \right)$. The evolution of the amplitudes $b_- $ and $b_+ $ of the $ e^{i \theta_-} |-\rangle$ and $e^{i \theta_+} |+\rangle$ states are driven by the hamiltonian $ \tilde H = R_\theta^\dag H R_\theta - i \hbar R_\theta^\dag \dot R_\theta$:
$$
\tilde H =   \frac{\hbar}{2} \begin{pmatrix} \omega_L \cos \theta - \omega_0  + 2 \dot \theta_- & -\Omega_L  e^{i \theta_{+ - } } \cr 
- \Omega_L^*  e^{i \theta_{- + }} &   \omega_0  - \omega_L \cos \theta + 2 \dot \theta_+
\end{pmatrix} 
$$
with $\theta_{+ -}=\theta_+-\theta_- = - \theta_{- +} $. 
One natural choice for $\theta_\pm $ is the ``first order'' choice 
 $\theta_\pm^{(1)} = \mp \frac{1}{2}\int^t_0 \omega_0 - \omega_L \cos \theta$  annulling the whole diagonal terms. 

Let us treat the (Schwinger 1937) example,  where all the parameters $\omega_0,\theta,\dot \varphi=-\omega_L$ are real and time independent.
The evolution operator in the adiabatic $|\mp\rangle$ basis $ \tilde U (t,0) = R_\theta^\dag (t) U (t,0) R_\theta (0)$ (where $U (t,0)$ is the evolution operator in the diabatic basis) verifies $i \hbar \dot{\tilde{U}}  = \tilde H \tilde U$ and, 
with $\theta_\pm = \theta_\pm^{(1)}$, is given
 by the matrix:
$$\tilde{U}=
\left( \begin{smallmatrix}
(\cos\frac{\Omega_R t}{2}-i\frac{\delta_L}{\Omega_R}\sin\frac{\Omega_R t}{2} ) e^{i \frac{\delta_L t}{2} } & i e^{i \frac{\delta_L t}{2} } \frac{\Omega_L}{\Omega_R}\sin\frac{\Omega_R t}{2} \\
i e^{-i \frac{\delta_L t}{2} } \frac{\Omega_L}{\Omega_R}\sin\frac{\Omega_R t}{2}
&( \cos\frac{\Omega_R t}{2}+i\frac{\delta_L}{\Omega_R}\sin\frac{\Omega_R t}{2} ) e^{-i \frac{\delta_L t}{2} } \end{smallmatrix} \right)
$$ 
The adiabaticity (negligible off-diagonal terms in $\tilde U$) evolution is given by 
the following condition
$
A^{(2)} =  \frac{|\Omega_L|}{\Omega_R} = \frac{| \omega_L \sin \theta|}{\sqrt{\omega_0^2-2 \omega_L \omega_0 \cos \theta + \omega_L^2 }}   \ll 1
$, where here $\Omega_R=  \sqrt{\Omega_0^2+\delta_0^2}  $ is
the generalized Rabi frequency. Using $\Omega_R = \sqrt{|\Omega_L|^2+\delta_L^2}$ this adiabatic condition can  be written 
\begin{equation}
2 A^{(1)} =  \left|  \frac{\Omega_L}{\delta_L} \right| = \left| \frac{ \omega_L \sin \theta}{\omega_0 - \omega_L \cos \theta } \right|  \ll 1
\label{cond_2_level}
\end{equation}
which has to be compared with
the
 ``usual'' adiabatic condition
 given by Eq. (\ref{adia_crit}): 
\begin{equation}
2 A^{(0)} = \left| \frac{ \Omega_L}{\omega_0} \right|= \left| \frac{ \omega_L \sin \theta  }{\omega_0} \right| \ll  1.
\label{usual_cond_2_level}
\end{equation}
$A^{(0)}, A^{(1)},A^{(2)}$ notations  will be generally defined latter.

Looking at the 
 $\omega_L \approx \omega_0$ and $\theta$ very small   resonant   case ($\delta_L\approx\delta_0 \approx  0  $), we see, in a simpler way than in Ref. \cite{2004PhRvL..93p0408M,2005PhRvL..95k0407T}
 and contrary to what is sometimes claimed  \cite{2005PhLA..339..288T,2004quant.ph..5129P}, that 
 Eq. (\ref{usual_cond_2_level}) is verified but not Eq. (\ref{cond_2_level}).  
This fundamental conclusion, based on a hamiltonian $H(t) \neq \hat H(t/T)$ is still valid for the time scaling case $\hat H (t/T)$. Indeed,
 the Schwinger hamiltonian can be of the $\hat H (t/T)$ type 
 if $\omega_L T$ is taken to be constant,
 for instance by looking at the evolution after one period $T=T_L = 2\pi/\omega_L$ depending on the $\omega_L$ parameter value.
 Indeed,
 $2A^{(1)} =  \frac{1}{T_L} \left| \frac{  \sin \theta  }{\omega_0} 
 + O(T_L^{-2}) \right| =  2A^{(0)} \left| 1+ \frac{2 A^{(0)}}{\tan \theta} + O(T_L^{-2})  \right|  $
  indicates, for instance if $\theta$ is very small,    why  
  an
evolution time $ T_L$ much longer  than expected by the usual condition ($ T_L \gg \left| \frac{  \sin \theta  }{\omega_0} \right| $ ) can be  needed to provide adiabatic evolution.
 \emph{The ``usual'' adiabatic conditions
 are then clearly not sufficient to provide adiabatic evolution even for $\hat H (t/T)$ hamiltonian type}.

 To be more general let us now study  a   discrete,  but possibly degenerate, hamiltonian with the state evolution
$|\Psi (t) \rangle = \sum_{m=1}^N b_m(t)  e^{i  \theta_m(t) }  |m (t)\rangle 
$ ($N \geq 2$). 
The phase 
$\theta_m =\gamma_m +  \alpha_m $ is real but not necessary
 equals to the first order choice
$\theta_m^{(1)} = \int_0^t i \langle m | \dot m \rangle - \int_0^t E_m/\hbar $: geometrical phase (which is the Berry Phase for cyclic evolution) plus dynamical  phase neither
contains the (Pancharatnam) phase $\arg \langle m(0) | m(t) \rangle $.
  To study the adiabatic evolution we shall assume that $| \Psi(t=0) \rangle = | n (0)\rangle$ (i.e. $b_n (0) =1$).
The evolution is adiabatic if 
$1-| \langle n (T) |  \Psi(T) \rangle | = 1- | b_n(T) | \ll 1 $  or equivalently  if $\|  |\Psi \rangle \langle \Psi |- | n \rangle \langle n | \| = \sqrt{1-   |b_n^2| } \ll 1 $  \cite{2006quant.ph..3175J}.

The Schr\"odinger's equation leads 
 for each $m$ state to:
\begin{equation}
 \dot b_m   = - i b_m \left( \dot \theta_m - \dot \theta_m^{(1)} \right) - \sum_{k \neq m}  b_k \langle m |  \dot k   \rangle  e^{i \theta_{k m} }  
  \label{eq_diff_b}
\end{equation}
 where $\theta_{ k m} = \theta_ k - \theta _m$.  
Using $\dot \theta_m = \dot \theta_m^{(1)}$,
 $ - \frac{ d |b_n| }{d t}   \leq  \left|  \frac{ d b_n }{d t} \right|$ and the norm inequality $\sqrt{N-1} \sqrt{1-   |b_n^2| } = \sqrt{N-1} \sqrt{ \sum_{m \neq n} | b_m|^2 } \geq \sum_{m \neq n}  |b_m| $  we find the first (very restrictive)  valid adiabatic condition for the interaction time $T$:
\begin{equation}
1-|b_n (T) |  \leq 1- \cos (\sqrt{N-1} \Omega_n  T) \leq (N-1) \frac{\Omega_n^2}{2} T^2  
\label{adi_cond_1}
\end{equation}
where
 $\Omega_n =
 \mathop{\max_{t \in [0,T]}}_{m \neq n}
 \left| \left \langle n(t) \left| \dot m(t)  \right. \right \rangle \right| $.
 This condition is  optimal  because it is reached  (see $\tilde U$) by the Schwinger $N=2$ level system for $\delta_L =0$ ($\Omega_n =|\Omega_L| = \Omega_R$). It illustrates the  quantum Zeno effect: during a time much smaller than $ \frac{1}{\sqrt{N} \Omega_n}$ the  system evolution is frozen.

In order to find  more useful adiabatic conditions
we 
integrate by part  Eq. (\ref{eq_diff_b})
using (for $k \neq m$)
$
A_{k m}  = \frac{\langle m | \dot k \rangle  e^{i (\theta_{k m }-\gamma_{k m })} 
e^{i ( \theta_k^{(1)} -\theta_k )}
}{  \dot \gamma_{k m} } $:
\begin{eqnarray}
\lefteqn{b_m(T)  - b_m (0) =  \sum_{k \neq m}   \left[ i b_k(t) e^{i  \gamma_{k m } (t) }
e^{i  (\theta_k (t)  -  \theta_k^{(1)} (t) ) }
   A_{ k m} (t)  \right]_{0}^{T} }  \nonumber  \\
   & &
- i \int_0^T  b_m \left( \dot \theta_m -  \dot \theta_m^{(1)} 
- \sum_{k \neq m}  
e^{i (\gamma_{k m}+\theta_{ m }   -  \theta_k^{(1)} ) } A_{k m}  \langle k |  \dot m   \rangle 
\right)
  \nonumber  \\
& &   
   - i  \sum_{k \neq m} \int_0^T      b_k   e^{i  \gamma_{k m } }  e^{i  (\theta_k -  \theta_k^{(1)}  ) } 
 \dot A_{ k m}  \label{eq_totale}  \\
 & &
 + i  \sum_{k \neq m} \int_0^T      b_k 
 \sum_{j \neq k,m}  e^{i (\gamma_{j m}+ \theta_k -  \theta_j^{(1)}  ) } A_{j m}  \langle j |  \dot k   \rangle  
 \nonumber 
\end{eqnarray}

It is now straightforward, with $m=n$, to look back to the standard adiabatic theorem with the time scaling $t = s T$. The evolution equation for $|\hat \Psi (s) \rangle = |\Psi (t(s)) \rangle$, 
is then $i \frac{\hbar}{T} \frac{d}{d s} |\hat \Psi (s) \rangle = \hat H (s) |\hat \Psi (s) \rangle$ and  the $T \rightarrow + \infty$ limit is similar to $\hbar \rightarrow 0$. 
With $\gamma_{k m } = E_{m n}/\hbar$, we have (for $\theta_k = \theta_k^{(1)}$) $
A_{k m}  = A_{k m}^{(0)}   =  \frac{
 \langle m | \dot k \rangle  e^{ -\int (\langle k | \dot k \rangle -\langle m | \dot m \rangle)}
  }{  
     \frac{E_m-E_k}{\hbar}  
  }
$ 
and
the stationary phase theorem 
(saddle-point or steepest descent method) annuls, for $T \rightarrow + \infty$,  the integrals in Eq. (\ref{eq_totale}) leading  to valid quantum adiabatic condition:
$$ 
 \sum_{m \neq n} \frac{1}{T} \left| \frac{ \left. \hbar \frac{d \hat H}{ d s} \right|_{ m n}  }{(E_{m n} )^2 } \right|  
 + o\left( \frac{1}{T} \right)  = \sum_{m \neq n} |A_{m n}^{(0)} | + o\left( \frac{1}{T} \right)  \ll 1   $$
A comparison with Eq. (\ref{adia_crit}) indicates, as also shown by the two level model where $|A_{+ -}^{(0)}|  = \frac{|\Omega_L|}{2 |\omega_0|}$, that a better understanding of the $o\left( \frac{1}{T} \right)$ term 
is in fact needed to have useful condition  \cite{2006quant.ph..3175J}.
 
 We could now go back to the general $H(t)$ case. 
$\sqrt{1- |b_n(T)|^2}$ verifies  $\sqrt{1- |b_n(T)|^2} \leq    \sqrt{N-1} b_-   $ with $b_- = \max_{t \in [0,T]} \max_{m \neq n} |b_m(t)|$. Using Eq. (\ref{eq_totale}) and $\theta_m = \theta_m^{(1)}$ choice, it be bounded by
$$
 b_-  \leq  \frac{ 2    A   +    \int  \! \!  |A'|   +   (N-2)  A   \Omega T }{ 1 -  (N-2) ( A +  \int    \! \!   | A' | ) - ((\mathtt{N-1} ) + (N-2)^2 ) A   \Omega T }.
$$
The typewriter style, such as $\mathtt{ ( N-1  ) A \Omega T} $, indicates terms that can be annulled by using  a better phase for $\theta_m$ namely the ``second order'' one
$\theta_m^{(2)} =
 \theta_m^{(1)} 
+ \int_0^t \sum_{k \neq m}  
e^{i ( \gamma_{k m}+\theta_m -\theta_k^{(1)} )} A_{k m}  \langle k |  \dot m   \rangle$.
The
three  important parameters are:
\begin{eqnarray*}
\Omega & = &  \mathop{\max_{t \in [0,T]}}_{k \neq m}   | \langle m |  \dot k   \rangle | =  \max_{m} \Omega_m
  \leq \max_{t \in [0,T]} \frac{\|\dot H \| }{ \Delta E  }  \\
A &= & A(T) = \mathop{\max_{t \in [0,T]}}_{k \neq m}  |A_{  k m }(t)| \leq \frac{\Omega}{ \displaystyle \mathop{\min_{t \in [0,T]}}_{k \neq m} \dot \gamma_{k  m} }  \\ 
 \int \! \! \! |A'| & = & \max_{k \neq m} \int_0^T \left| \dot A_{ k m}   \right|  
 \end{eqnarray*} 
Where, $\Delta E = \min_{k \neq m} E_{k m}$ is the energy spectrum gap. Another (better for large $T$) bound for $b_+ (T) = \min_{t \in [0,T]}  |b_n(t)| = |b_n(t_T)| $
is
obtained using $m=n$ in Eq. (\ref{eq_totale}) and the norm inequality: 
\begin{eqnarray*}
\lefteqn{ 1-  |b_n(t_T)|  \leq  \mathtt{ ( N-1 )  A   \Omega T } } \\ 
& & + \sqrt{N-1} \sqrt{1-b_+^2} (A   +    \sqrt{N-1}  \int   \! \! \!  |A'|   +   (N-2)  A   \Omega T )
\end{eqnarray*}
and a point fix study leads to
\begin{eqnarray}
\lefteqn{
1- b_+  \leq   \mathtt{  2 (N-1)  A   \Omega T } }  \label{bound_bn2}  \\
&  & + 2 (N-1)  (A   +    \sqrt{N-1}  \int  \! \! \!  |A'|   +   (N-2)  A   \Omega T  )^2 . \nonumber
\end{eqnarray}
Finally  one (not optimized) adiabatic condition is 
\begin{eqnarray}
   A    +   \sqrt{N}   \! \!  \int  \! \! \!  |A'| + ( \sqrt{ \mathtt{ N } } + N-2)  A \Omega T \ll \frac{1}{\sqrt{N}} 
     \label{adi_cond_3} 
\end{eqnarray}

We  define two useful \emph{reals} $A_{k m }$: 
 $$
A_{ k m}^{(1)}  =  \frac{ | \langle m | \dot k \rangle | }{    i (\langle k | \dot k \rangle -\langle m | \dot m \rangle) -  \frac{E_k-E_m}{\hbar} + \frac{d  } { d t} \arg \langle m | \dot k \rangle  } = \frac{ | \langle m | \dot k \rangle | }{   \dot \gamma_{k m }^{(1)}}
$$
 for the  $\theta_{k m} = \theta_{k m}^{(1)}$ choice ,
 and $A_{ k m}^{(2)}  = \frac{ | \langle m | \dot k \rangle | }{   \dot \gamma_{k m }^{(2)}}  $  for the
 $\theta_{k m} = \theta_{k m}^{(2)}$ choice where $\dot \gamma_{k m }^{(2)}   =   \dot \gamma_{k m}^{(1)} + \sum_{j\neq m}
\frac{|\langle \dot m | j  \rangle|^2 }{\dot \gamma _{j m }^{(2)}  } $.
    When  the  hamiltonian $H(t)$ is real  in the canonical basis, the eigenstates $| m \rangle$ and $\langle m | \dot k \rangle$ are reals and
   $\langle m | \dot m \rangle =0$ so, $  |A_{k m}^{(1)}| =   |A_{k m}^{(0)}|$. 
  
If all $A_{ k m}^{(1)}$, or $A_{ k m}^{(2)}$, are \emph{monotonics} in $[0,T]$
 $
\int_{0}^{T} |\dot A_{k m} |  = |A_{k m} (T) - A_{k m} (0)| $ and
 the  condition (\ref{adi_cond_3}) becomes simpler: 
$
      A^{(1)}    +     \sqrt{N} A^{(1)} \Omega T \ll 1/N $   or  $
          A^{(2)}  +  \sqrt{N-2} A^{(2)} \Omega T \ll 1/N$, where $ A^{(i)} $ indicates that it should be calculated using the $ A_{k m}^{(i)} $ choice.   
For $N=2$
 smallness and monotonicity  of 
 $A_{+ -}^{(1)} =  \frac{|\Omega_L|}{2 \delta_L}$ 
 is equivalent to smallness and no more than one monotonicity change of $A_{+ -}^{(2)} =  \frac{|\Omega_L|}{\delta_L +\sqrt{\delta_L^2 + |\Omega_L|^2} } \geq 0 $. Thus, a final general, simple and useful adiabatic condition is  (for monotonics $A^{(1)}_{k m}$)
\begin{equation}
     A^{(1)}  +  \sqrt{N-2} A^{(1)} \Omega T \ll 1/N. \label{adi_cond_4}
\end{equation}
It is even possible to refine the condition by  dividing the interval $[0,T]$ in smaller intervals where all $ A_{k m}^{(i)} $ 
are monotonics. A perturbative point of view, neglecting  the $A^{(1)} \Omega T$ term,  has been used to derive similar results \cite{2005quant.ph..9083Y}.

The 
$N=2$ case is illustrative 
because 
it is the only one where a
 time independent adiabatic condition exists:
\begin{equation}
 2 |A^{(1)}| = \left|  \frac{\Omega_L}{\delta_L } \right| \ll \frac{1}{M^2} 
\label{two_level_cond}
\end{equation}
 where $M-1$ is the number  of monotonicity change of $ \frac{|\Omega_L|}{ \delta_L} $ in $[0,\infty]$. 
 This generalize the   Schwinger conditions
 Eq. (\ref{cond_2_level}). 
 For real hamiltonian the condition is
 $$2|A^{(1)}| =  2|A^{(0)}| = \left| \frac{\dot \theta}{\omega_0} \right|  = \left| \frac{\Omega_0 \dot \delta_0 }{(\delta_0^2 + \Omega_0^2 )^{3/2}} \right| \ll  \frac{1}{M^2} $$ 
 and becomes the usual adiabatic condition  if $M$ is small, for instance 
if the  matrix elements    $    \dot \delta_0,  \Omega_0 $ of $H$ and $\dot H$
  have 
  small number of monotonicity changes. 
    This explain why
the real
  dressed state hamiltonian,
 $H_0 = R_0^\dag H R_0 - i \hbar R_0^\dag \dot R_0= - \frac{\hbar}{2}
 \left( \begin{smallmatrix}
\delta_0  & \Omega_0 \cr 
 \Omega_0 & - \delta_0\end{smallmatrix} \right)$,
  obtained from $H$ 
 in the  rotating frame (with the simple phase choice $\theta_+=\theta_-=0$) or simply by $\omega_L=0$,
have been \emph{luckily}
combined with the usual adiabatic theorem
 to describe several adiabatic evolutions such as, the RAP
 (Rapid Adiabatic Passage), the  SCRAP (frequency or Stark-Chirped RAP) or  the STIRAP (STImulated Raman Adiabatic Passage).
  
  However
     when  real oscillatory terms are present the usual adiabatic  condition is no more sufficient to provide adiabatic evolution.
 As
 example we use the cycling hamiltonian \cite{Grifoni,2005JPhA...38.9979M}, $H= H_0$
 with $\delta_0(t) =  \alpha \cos (\omega t)$ and $\alpha,\omega,\Omega_0$ are (positives to simplify) constants. It
is relevant in many areas in physics: magnetic resonance, atomic collision, laser-atom interactions without the RWA and even localization by
exchanging the parameters $\delta_0$ and $\Omega_0$ (hamiltonian $ R_y   H_0 R_y^\dag $ with $R_y=e^{i \pi \sigma_y/4}$).
The weak-coupling and large amplitude case $\alpha \gg \Omega_0, \omega$ is simple  because the non-adiabatic transition probability $p_1$
(so called single-passage or one-way  transition) 
 is given by one of
the simplest of the
  several existing approximate formulas
  (Landau-Zener-St\"uckelberg, Rosen-Zener-Demkov, Nikitin, Zhu-Nakamura models, ...
 \cite{Nakamura2002,Nikitin2006}) namely the
Landau-Zener  one: $p_1 \approx e^{-2 \pi \frac{\Omega_0^2}{4 \alpha \omega} } = e^{-  \pi /( 4 A^{(1)} (\infty) ) } $
\cite{1994PhRvA..50..843K}.  
The $M=2$ double-passage transition probability $p_2$, which depends of a relative  (St\"uckelberg) phase $\Theta$ 
 of the wavefunction, 
   $p_2 = 4 p_1 \sin^2 (\Theta) $  can  be
   $4$ times higher than $p_1$ 
 and
 the  $M$ (even) multiple passage probability 
   $p_{M} \approx p_1 \frac{\sin^2 M\Theta}{\cos^2 \Theta} $ can be
 $M^2$ times higher than $p_1$. Here small $\omega$ value leads
to the adiabatic limit $p_1\rightarrow 0$ and with  $ \Theta \simeq \frac{\alpha}{\omega} \sim \pi/2$ we could have
$p_{M} \sim 1$ \cite{1994PhRvA..50..843K}.
Interestingly enough, 
the reverse case, namely the diabatic limit ($p_1\rightarrow 1$) can leads (for instance when  $ \alpha/\omega$ annul the Bessel $J_0$ function) to the reverse phenomenum of adiabaticity created after  multiple passages ($p_M \approx 0$) 
known as suppression of the tunneling,  coherent destruction of tunneling, 
 dynamical localization or  population trapping depending on the context   
\cite{Grifoni,1994PhRvA..50..843K}. 

This two level example illustrate why  monotonicity is require to avoid  constructive  interferences transforming  an adiabatic (resp. diabatic) single passage 
in a  fully diabatic (resp. adiabatic)  transition after multiple passages.
The two level system with several crossings is very similar to the case of single crossing but with several levels leading to sum of
dephased Landau-Dykhne-Davis-Pechukas  formulas 
\cite{1991PhRvA..44.4280J,2004AcPPB..35..551G}. Moreover, the
transition probability
 in a multilevel system is the product of several
 Landau-Dykhne type terms, corresponding to several successive
transitions between pairs of levels \cite{2000PhRvA..61f2104W}. 
However, several consecutive  constructive interferences are exceptional and the generic most common case concern a system
 ``complex enough'' with small total probability  when the single crossing probability is small \cite{Akulin2006}.

 In conclusion,  we have derived exact bounds for the evolution Eqs. (\ref{adi_cond_1}), (\ref{bound_bn2}) as well as general adiabaticity criterion Eqs. (\ref{adi_cond_4}), (\ref{two_level_cond}).
 The key parameters for adiabaticity are the smallness and the small number of monotonicity change   of 
$ A^{(1)} \sim \frac{1}{\gamma^{(1)}}  \frac{ \| \dot H \| }{\Delta E  }  $ as well as a short  evolution time ($T^{-1} \gg (N-2)^{3/2} \frac{1}{\gamma^{(1)}}  \frac{ \| \dot H \|^2 }{{\Delta E}^2  } $).
For real hamitonian the adiabatic (Pancharatnam) phase type $\gamma^{(1)}$ is the spectrum frequency gap and  the usual  adiabatic condition are restored 
if the  matrix elements  of $H$ and $\dot H$
  have 
  small number of monotonicity changes in the two level ($N=2$) case.
 The results presented here, and demonstrated for the discrete, but possibly degenerate case, might be  useful for adiabatic quantum evolution and adiabatic quantum computation studies. 
  Extension  to the infinite dimensional or non hermitian cases are some of the next steps needed  to derive more universal quantum adiabatic conditions. 
  
  The author acknowledge Andr\'ea Fioretti for helpful discussions.
  
 This work has been realized in the framework of "Institut francilien de recherche sur les atomes
froids" (IFRAF).

\end{document}